\documentstyle[12pt]{article}
\long\def\del#1\enddel{ }
\let\3=\ss \catcode`\"=\active \let"=\"
  \oddsidemargin=\evensidemargin
\let\ni=\noindent  
 \let\msk=\medskip 
  
\let\qd=\quad \let\qqd=\qquad  

\let\a=\alpha \let\b=\beta \let\g=\gamma \let\d=\delta \let\e=\varepsilon
    
\let\l=\lambda \let\m=\mu \let\n=\nu \let\x=\xi \let\p=\pi 
\let\s=\sigma \let\t=\tau  
\let\ps=\psi 
     
 \let\L=\Lambda  \let\D=\Delta
\def\0{\over }    \def\1{\vec }   \def\2{{1\over2}} \def\3{{\ss}}
\def\4{{1\over4}} \def\5{\bar }   \def\6{\partial } \def\7#1{{#1}\llap{/}}
\def\8#1{{\textstyle{#1}}}        \def\9#1{{\bf {#1}}}
\def\_#1{$\underline{\hbox{#1}}$} \def\^#1{$\overline{\hbox{#1}}$}

\def\<{\langle } \def\>{\rangle }  
 
\def \({\left( } \def \){\right) }

    \let\hc=\dagger 
      \let\and=\wedge
     
\def\|#1{{}_{\bigg|_{#1}}}

\def\mao#1{\mathop{\rm {#1}}\nolimits}     \def\Tr{\mao{Tr}}
  \def\Im{\mao{Im}}

\def\beq{\begin{equation}} \def\eeq{\end{equation}} \def\eql#1{\label{#1}\eeq}
\def\bea{\begin{eqnarray}} \def\eea{\end{eqnarray}} \def\eal#1{\label{#1}\eea}
\let\nn=\nonumber

\def\plb#1 #2 {Phys. Lett. {\bf B#1} #2 }
\def\phr#1 #2 {Phys. Rep. {\bf  #1} #2 } 
\def\npb#1 #2 {Nucl. Phys. {\bf B#1} #2 }
\def\aph#1 #2 {Ann. Phys. {\bf #1} #2 }  
\def\jmp#1 #2 {J. Math. Phys. {\bf #1} #2 }
\def\prd#1 #2 {Phys. Rev. {\bf D#1} #2 }
\def\prl#1 #2 {Phys. Rev. Lett. {\bf #1} #2 }
\def\rmp#1 #2 {Rev. Mod. Phys.  {\bf #1} #2 }
\def\zpc#1 #2 {Z. Phys. {\bf #1C} #2 }
\def\cmp#1 #2 {Comm. Math. Phys. {\bf #1} #2 }
\def\mpl#1 #2 {Mod. Phys. Lett. {\bf A#1} #2 }
\def\ijmp#1 #2 {Int. J. Mod. Phys. {\bf A#1} #2 }
\def\jpa#1 #2 {J. Phys. {\bf A#1} #2 }

\begin{document}
{\hfill ITP--UH--10/94\vskip-9pt     \hfill hep-th/9407086}
\vskip 30mm \centerline{\hss\Large \bf
    RENORMALIZATION GROUP FLOW \hss}
\vskip 7mm\centerline{\hss\Large \bf
    IN A GENERAL GAUGE THEORY  \hss}
\begin{center} \vskip 12mm
       {\large Harald SKARKE}
\vskip 5mm
       Institut f"ur Theoretische Physik,
       Universit"at Hannover\\
       Appelstra\3e 2,
       D--30167 Hannover,
       GERMANY
\vskip 5mm
      e-mail: skarke@itp.uni-hannover.de
\vfil                        {\bf ABSTRACT}                \end{center}

The renormalization group flow in a general renormalizable gauge theory
with a simple gauge group in 3+1 dimensions is analyzed.
The flow of the ratios of the Yukawa couplings and the gauge coupling
is described in terms of a bounded potential, which makes it possible
to draw a number of non-trivial conclusions concerning the asymptotic
structure of the theory.
A classification of possible flow patterns is given.

\vfil\noindent \\ July 1994 \msk
\thispagestyle{empty} \newpage
\setcounter{page}{1} \pagestyle{plain}

\section{Beta functions of a general gauge theory}
When mass effects can be neglected, the behaviour of a quantum field
theory under scale transformations can be described conveniently in terms of
``running couplings'' defined via beta functions.
Whereas it is quite straightforward to calculate the beta functions, it is
usually not so simple to deduce from
them the structure of the renormalization group flow, especially when there
is a large number of coupling constants.
In the present work we will show that even for a very general theory it
is possible to make a number of nontrivial statements concerning its
asymptotic behaviour.

We consider a gauge theory with a simple gauge group $G$ and with
Weyl fermions $\ps_i$ and real scalars $\phi_\a$ transforming under some
(generically reducible) representations $R_F$ and $R_B$ of $G$.
The Lagrangian of this model is
\beq 
-{1\04}F^a_{\m\n}F^{a\m\n}
+i\bar\psi_i\bar\s^\m D_\m\psi_i+\2D_\m\phi_\a D^\m\phi_\a
-\2\psi_i\psi_j Y^\a_{ij}\phi_\a-\2\bar\psi_i\bar\psi_j {Y^\a_{ij}}^*\phi_\a
-{1\0 24}V^{\a\b\g\d}\phi_{\a}\phi_{\b}\phi_{\g}\phi_{\d},     \eql{lag}
where we have omitted gauge fixing, ghost, $\phi^3$ and mass terms.
Gauge invariance forces the Yukawa couplings $Y_{ij}^\a$ and the quartic
scalar couplings $V^{\a\b\g\d}$ to fulfill the relations
\beq Y_{kj}^\a(T_F)^a_{ki}+Y_{ik}^\a(T_F)^a_{kj}+Y_{ij}^\b(T_B)^{a\b\a}=0
                                                              \eql{niy}
and
\beq V^{\e\b\g\d}(T_B)^{a\e\a}+V^{\a\e\g\d}(T_B)^{a\e\b}+
     V^{\a\b\e\d}(T_B)^{a\e\g}+V^{\a\b\g\e}(T_B)^{a\e\d}=0,       \eql{niv}
respectively. $T_F$ and $T_B$ are the generators of $R_F$ and $R_B$.
These equations could be solved by decomposing the group representations
according to $\psi_i\to\psi^I_{\tilde i}$, $\phi_\a\to\phi_{\tilde\a}^A$,
where $I,A$ run through the sets of irreducible representations
spanned by $\tilde i=\tilde i(I)$ and $\tilde \a=\tilde \a(A)$. Then the
couplings can be written as
\beq Y^\a_{ij}\to Y^{\tilde\a IJ}_{A\,\tilde i\,\tilde j}=
     \sum_k(Z^{(k)})^A_{IJ}
           (\L^{(k)})^{\tilde\a}_{\tilde i\tilde j},      \eql{dec}
where the $\L^{(k)}$ are tensors satisfying an analogue of Eq. (\ref{niy})
whereas the $(Z^{(k)})^A_{IJ}$ are the usual unrestricted coupling constants.
In the same way also the $V^{\a\b\g\d}$ can be decomposed.
It will turn out, however, that for a general analysis it is more convenient
not to make use of this decomposition.

The one-- and two--loop contributions to the beta functions
of this model have been calculated in the $R_\x$ gauge in dimensional
regularization with the $\overline{MS}$ scheme \cite{calc}.
The gauge beta function is given by
\bea \b(g)=\m{dg\0d\m}&=&
   -{g^3\0 6(4\p)^2}(22c_g-4S_F-S_B)
   -{g^3\0 (4\p)^4d_g}\Tr(C_F{Y^\b}^\hc Y^\b)
   \nn\\ &&+{g^5\0 3(4\p)^4}[6Q_F+6Q_B+c_g(10S_F+S_B-34c_g)]+O(g^7).  \eal{bg}
$C=T^aT^a$ is the quadratic Casimir Operator, $d_g$ the dimension of $G$,
$S={1\0d_g}\Tr C$ the Dynkin index, and $Q={1\0d_g}\Tr C^2$.
The counting of orders in $g$ is such that $Y=O(g)$, $V=O(g^2)$.
The other beta functions are
\bea \b(Y^\a)=\m{dY^\a\0d\m}=
{1\0 2(4\p)^2}&&[4Y^\b{Y^\a}^\hc Y^\b+Y^\a{Y^\b}^\hc Y^\b+Y^\b{Y^\b}^\hc Y^\a
\nn\\ && +Y^\b\Tr({Y^\a}^\hc Y^\b+{Y^\b}^\hc Y^\a)-6g^2(Y^\a C_F+C_FY^\a)]
    + O(g^5)    \eal{Yuk}
(using matrix notation for the fermionic indices) and
\bea &&(4\p)^2\b(V^{\a\b\g\d})=
     V^{\a\b\l\e}V^{\g\d\l\e}+V^{\a\g\l\e}V^{\b\d\l\e}+
     V^{\a\d\l\e}V^{\b\g\l\e}\nn\\
     &&\qqd+[\2\Tr({Y^\a}^\hc Y^\l+{Y^\l}^\hc Y^\a)-3g^2C_B^{\a\l}]V^{\l\b\g\d}
       \nn\\
     &&\qqd+[\2\Tr({Y^\b}^\hc Y^\l+{Y^\l}^\hc Y^\b)-3g^2C_B^{\b\l}]V^{\a\l\g\d}
       \nn\\
     &&\qqd+[\2\Tr({Y^\g}^\hc Y^\l+{Y^\l}^\hc Y^\g)-3g^2C_B^{\g\l}]V^{\a\b\l\d}
       \nn\\
     &&\qqd+[\2\Tr({Y^\d}^\hc Y^\l+{Y^\l}^\hc Y^\d)-3g^2C_B^{\d\l}]V^{\a\b\g\l}
    \nn\\
   &&\qqd+3g^4(\{T^a_B,T^b_B\}^{\a\b}\{T^a_B,T^b_B\}^{\g\d}+
           \{T^a_B,T^b_B\}^{\a\g}\{T^a_B,T^b_B\}^{\b\d}+
           \{T^a_B,T^b_B\}^{\a\d}\{T^a_B,T^b_B\}^{\b\g})\nn\\
&&\qqd-2\Tr[({Y^\a}^\hc Y^\b+{Y^\b}^\hc Y^\a)({Y^\g}^\hc Y^\d+{Y^\d}^\hc Y^\g)
       \nn\\
&&\qqd\qqd+({Y^\a}^\hc Y^\g+{Y^\g}^\hc Y^\a)({Y^\b}^\hc Y^\d+{Y^\d}^\hc Y^\b)
       \nn\\
&&\qqd\qqd+({Y^\a}^\hc Y^\d+{Y^\d}^\hc Y^\a)({Y^\b}^\hc Y^\g+{Y^\g}^\hc Y^\b)]
     +O(g^6).                                                      \eal{ph4}
In order to solve the system of ordinary differential equations (\ref{bg})
-- (\ref{ph4}) it is convenient to change variables to
$t=\ln(\m/\m_0)$, $G=(g/4\p)^2$, $y=Y/g$ and $v=V/g^2$.
Then the evolution of $G$ is governed by
\beq {dG \0 dt}= -\l G^2-{2G^3\0 3}\(34c_g^2-10c_gS_F-c_gS_B-6Q_F-6Q_B+
               {3\0d_g}\Tr(C_F{y^\b}^\hc y^\b)\)+O(G^4),  \eql{bG}
where
\beq \l=(22c_g-4S_F-S_B)/3.   \eeq
To lowest order the solution is given by
\beq G^{-1}(t)=G^{-1}_0+\l t+O(G), \eeq
where $G_0=G(t=0)$ (similar notation will be used again).
$y$ and $v$ evolve according to
\beq {dy^\a\0dt}=
      {G\0 2}(4y^\b{y^\a}^\hc y^\b+y^\a{y^\b}^\hc y^\b+y^\b{y^\b}^\hc y^\a
      +y^\b\Tr({y^\a}^\hc y^\b+{y^\b}^\hc y^\a)-y^\a D-Dy^\a)
      +O(G^2),   \eql{yuk}
where
\beq D_{ij}=6(C_F)_{ij}-\l\d_{ij}/2, \eeq
and
\bea &&G^{-1}{dv^{\a\b\g\d}\0dt}=
     v^{\a\b\l\e}v^{\g\d\l\e}+v^{\a\g\l\e}v^{\b\d\l\e}+
     v^{\a\d\l\e}v^{\b\g\l\e}\nn\\
     &&\qqd+[\2\Tr({y^\a}^\hc y^\l+{y^\l}^\hc y^\a)-3C_B^{\a\l}+
         {\l\0 4}\d^{\a\l}]v^{\l\b\g\d} \nn\\
       &&\qqd+[\2\Tr({y^\b}^\hc y^\l+{y^\l}^\hc y^\b)-3C_B^{\b\l}+
         {\l\0 4}\d^{\b\l}]v^{\a\l\g\d}       \nn\\
     &&\qqd+[\2\Tr({y^\g}^\hc y^\l+{y^\l}^\hc y^\g)-3C_B^{\g\l}+
         {\l\0 4}\d^{\g\l}]v^{\a\b\l\d}\nn\\
       &&\qqd+[\2\Tr({y^\d}^\hc y^\l+{y^\l}^\hc y^\d)-3C_B^{\d\l}+
         {\l\0 4}\d^{\d\l}]v^{\a\b\g\l}    \nn\\
   &&\qqd+3(\{T^a_B,T^b_B\}^{\a\b}\{T^a_B,T^b_B\}^{\g\d}+
           \{T^a_B,T^b_B\}^{\a\g}\{T^a_B,T^b_B\}^{\b\d}+
           \{T^a_B,T^b_B\}^{\a\d}\{T^a_B,T^b_B\}^{\b\g})\nn\\
&&\qqd-2\Tr[({y^\a}^\hc y^\b+{y^\b}^\hc y^\a)({y^\g}^\hc y^\d+{y^\d}^\hc y^\g)
       \nn\\
 &&\qqd\qqd+({y^\a}^\hc y^\g+{y^\g}^\hc y^\a)({y^\b}^\hc y^\d+{y^\d}^\hc y^\b)
       \nn\\
 &&\qqd\qqd+({y^\a}^\hc y^\d+{y^\d}^\hc y^\a)({y^\b}^\hc y^\g+{y^\g}^\hc y^\b)]
     +O(G) ,                                                        \eal{bv}
respectively.
The forms of these equations suggest to define a new evolution
parameter $\t$ by
\beq d\t={G\0 2}dt\qd {\rm and}\qd \t_0=0.   \eeq
If $\l\ne 0$, then
\beq \t=-{1\0 2\l}\ln (G/G_0)+O(G\ln G), \eeq
whereas for $\l=0$
\beq \t={Gt\0 2}+O(G^3). \eeq

\section{Potentials}
In the center of our further discussion of the flow of $y$ there will
be a ``potential'' $U$ defined by
\beq U={1\0 3}\sum_{ijkl}|S_{ijkl}|^2
   +\sum_{\a\b}(I^{\a\b})^2+\sum_{ij}|M_{ij}|^2 \eql{U}
with
\bea S_{ijkl}&=&y^\a_{ij}y^\a_{kl}+y^\a_{ik}y^\a_{jl}+y^\a_{il}y^\a_{jk},\\
   I^{\a\b}&=&\Im\Tr{y^\a}^\hc y^\b,\\
{\rm and}\qd M&=&{y^\a}^\hc y^\a-D. \eea
It can easily be checked that, at lowest order in $G$,
\beq {2\0 G}{dy^\a_{ij}\0dt}={dy^\a_{ij}\0d\t}={\6 U\0\6{y^\a_{ij}}^\hc}, \eeq
implying
\beq {dU\0d\t}={\6 U\0\6{y^\a_{ij}}^\hc}{{dy^\a_{ij}}^\hc\0d\t}+
      {\6 U\0\6 y^\a_{ij}}{dy^\a_{ij}\0d\t}=2|{\6 U\0\6 y^\a_{ij}}|^2\ge 0\eeq
Thus $U$ increases or decreases with increasing or decreasing $\t$, along
paths of steepest ascent or descent, respectively.
$U$ plays a role similar to that of $c$ in Zamolodchikov's
$c$-theorem \cite{zam}.
In fact, potentials for the renormalization group flow have been considered
already 20 years ago \cite{wz}.
In general one has to define a metric with respect to which a gradient
flow is defined. In our particular case it turns out to be just the
Euclidean metric on $y$-space.
The sign of $U$ is chosen in such a way that $\m,t,\t,U$ are
monotonically increasing functions of one another.

{}From the explicit form of Eq. (\ref{U}) it is obvious that $U$ is
non-negative, and it is not hard to show that the entries of
$M$ become large when one of the $y^\a_{ij}$ becomes large.
Therefore $U$ has a global minimum, i.e. an infrared fixed point, at some
finite value of $y$.
If $R_F$ or $R_B$ contains isomorphic irreducible representations more than
once, Eqs. (\ref{yuk}) and (\ref{bv}) are invariant under
some global symmetries relating these irreducible representations and
the set of fixed points generically will not be a unique point, but rather
the orbit of such a symmetry.
In fact it is also possible that the space of minima of $U$ has a
degeneracy that is not related to a global symmetry.
An example is given by the model of Ref. \cite{tf}, where a continuous set
of solutions of $U=0$ was found.
Since the eigenvalues of $\Tr{y^\a}^\hc y^\b$ would be invariant under
global symmetries, solutions with different eigenvalues
that were found cannot be related by symmetries.
It would be interesting to find out whether local minima at different values
of $U$ are possible and whether the space of global minima is connected.

It will also be important how fast the minimum is approached in the limit
$\t\to \infty$.
If the Hessian $H$ of $U$ at $y_{\rm FP}$ has maximal rank (i.e. if
$y_{\rm FP}$ is a non-degenerate critical point), all components of
$\D y=y-y_{\rm FP}$ will tend to zero exponentially in $\t$ because of
\beq {d\D y\0 d\t}=H\D y+O((\D y)^2). \eql{hess}
As we have seen, the minimum often has flat directions, implying that
the Hessian cannot have maximal rank.
If the minimum locus is locally a submanifold of $y$-space (this is
not fulfilled, for example, for the origin for $U=y_1^2y_2^2$), and if the
rank of the Hessian is the maximal rank $N$ (the dimension of $y$-space)
minus the number $N_{\rm flat}$
of flat directions, one can locally define coordinates by the following
procedure: Define $N_{\rm flat}$ coordinates parametrizing the minimum locus
and label each point in some suitably chosen open set
by the $N_{\rm flat}$ coordinates of the fixed point to which it flows and
by $N-N_{\rm flat}$ linear combinations of the original coordinates
on which the Hessian is positive definite.
Under the flow, the first $N_{\rm flat}$ coordinates do not change,
whereas the other $N-N_{\rm flat}$ coordinates approach the fixed point
exponentially in $\t$. Thus $||\D y||$ is again exponentially bounded.
If $N_{\rm flat}+$rank$(H)<N$, we cannot apply these arguments.
For example, if $d\D y/d\t\sim const.\times (\D y)^3$, then
$\D y\sim const.\times (const.-\t)^{-1/2}$.

We have written $U$ in terms of three highly symmetric structures which were
discovered in Ref. \cite{k} (see also \cite{kk}) in the context of a search
for finite theories.
A case of
special interest occurs when each of these structures vanishes separately.
In particular this is the case for one loop finite supersymmetric theories
\cite{k,k93}. For any supersymmetric theory $S_{ijkl}$ will vanish:
The Yukawa couplings come in two types, namely interactions involving gauginos,
which are proportional to generators of the gauge group, and interactions
within the chiral sector, which are determined by totally symmetric
constants $d_{ijk}$.
Within one type of couplings contributions corresponding to the real and
imaginary parts of the scalars cancel, whereas
the mixed terms vanish because of the invariance condition on $d_{ijk}$
(their contribution to $S_{ijkl}$ is just the analogue of the l.h.s. of Eq.
(\ref{niy})).
The conditions $I=0$ and $M=0$ are not fulfilled automatically and
lead to the well--known one-loop finiteness conditions for supersymmetric
theories \cite{ref2}.
An example for a non-supersymmetric theory with simultaneous vanishing of
$S_{ijkl}$, $I^{\a\b}$ and $M_{ij}$ is given in Ref. \cite{tf}.
There are many models, however, where it is not possible to solve $U=0$
\cite{gran}.

It is easy to see that $U$ has at most one local maximum:
Consider the one parameter family of couplings
$y^\a_{ij}(z)=z(y_0)^\a_{ij}$ with a fixed set of values $(y_0)^\a_{ij}$.
Then $U(y^\a_{ij}(z))$ is of the form $a|z|^4+b|z|^2+c$ with $a>0$, which
has at most one local maximum, namely at $z=0$. Therefore any maximum of $U$
can only be located at $y^\a_{ij}=0$ for all $\a,i,j$.
Clearly $y=0$ is always a critical point.
The Hessian there is diagonal (in the $d_B\times d_F^2$-dimensional $y$-space)
with diagonal elements corresponding to $y^\a_{ij}$ given by
$-(D_{ii}+D_{jj})$.
This point really represents a maximum if and only if the sub-matrix of
$D$ corresponding to Yukawa--interacting fermions is positive definite.
If this is the case, $y=0$ is approached according to
\beq y^\a_{ij}\sim const. \times e^{-(D_{ii}+D_{jj})\t}\eeq
(no summation over repeated indices).

In addition to this maximum and the set of minima, $U$ might (and usually will)
also have saddle points.
Particular examples are points where the Yukawa couplings for some of the
particles vanish while the Yukawa couplings for the other particles correspond
to some minimum of the restricted $U$.
Although saddle points are unstable fixed points of the flow both in the
infrared and in the ultraviolet limit, getting close to such a point (and
thereby getting small beta functions) can considerably decelerate the flow,
such that a realistic theory will often be close to a saddle point along large
portions of the flow.
A theory flowing exactly into a saddle point would require exactly
fine-tuned initial conditions. In the absence of reasons for such a
fine-tuning, this will occur with zero probability.
A possible reason might be a global symmetry which could impose
constraints on the couplings.
In such a case one should consider the potential $U$ only
over the subspace of $y$-space defined by these constraints.
Then the couplings run to extrema of the constrained potential.
A typical example is supersymmetry which prevents $y$ from running
to 0 in the ultraviolet limit.

Let us now discuss the behavior of $v$:
We consider Eq. (\ref{bv}) for fixed $y=y_{\rm FP}$, neglecting higher
orders in $G$.
It is possible to integrate to get a potential again:
\bea U_v&=&
     v^{\a\b\g\d}v^{\g\d\l\e}v^{\l\e\a\b}
     +2[\2\Tr({y^\a}^\hc y^\l+{y^\l}^\hc y^\a)-3C_B^{\a\l}+
         {\l\0 4}\d^{\a\l}]v^{\a\b\g\d}v^{\l\b\g\d} \nn\\
   &&\qqd+[3(\{T^a_B,T^b_B\}^{\a\b}\{T^a_B,T^b_B\}^{\g\d}+
           \{T^a_B,T^b_B\}^{\a\g}\{T^a_B,T^b_B\}^{\b\d}+
           \{T^a_B,T^b_B\}^{\a\d}\{T^a_B,T^b_B\}^{\b\g})\nn\\
&&\qqd-2\Tr(({y^\a}^\hc y^\b+{y^\b}^\hc y^\a)({y^\g}^\hc y^\d+{y^\d}^\hc y^\g)
    +({y^\a}^\hc y^\g+{y^\g}^\hc y^\a)({y^\b}^\hc y^\d+{y^\d}^\hc y^\b)  \nn\\
&&\qqd\qqd+({y^\a}^\hc y^\d+{y^\d}^\hc y^\a)({y^\b}^\hc y^\g+{y^\g}^\hc y^\b))]
    v^{\a\b\g\d}                                                   \eal{Uv}
In contrast to the potential for the Yukawa couplings, $U_v$ is unbounded,
because it is cubic.
We can show, however, that for any given set of $y$'s it has at most a single
local minimum:
Assume there are two minima $v_1, v_2$ and consider the line
$v(s)=v_1+s\D v$ with $\D v=v_2-v_1$.
Then $s=0$ and $s=1$ are minima of the function  $U_v(v(s))$ which is
(at most) cubic in $s$.
Therefore $U_v$  must be constant along $v(s)$.
Now consider $v(s)$ near $s=0$.
If every neighborhood of $s=0$ contains points that are not minima of (the
full)
$U_v$, then every neighborhood of $v_1$ (in the full $v$-space) will contain
points $v'$ with $U(v')<U(v_1)$, in contradiction to the assumption that
$v_1$ is a local minimum of $U_v$.
If, on the other hand, there is a neighborhood of $s=0$ containing only
minima of $U_v$, then $d(v_1+s\D v)/d\t$ must vanish identically.
The coefficient of $s^2$ in $d(v_1+s\D v)/d\t$ is proportional to
$\D v^{\a\b\l\e} \D v^{\g\d\l\e}+\D v^{\a\g\l\e} \D v^{\b\d\l\e}+
\D v^{\a\d\l\e} \D v^{\b\g\l\e}$.
Summation over $\a=\g$ and $\b=\d$ gives
\beq 2\sum_{\a\b\g\d}(\D v^{\a\b\g\d})^2+\sum_{\g\d}(\sum_\a\D v^{\a\a\g\d})^2
     =0, \eeq
implying $\D v=0$, i.e. $v_1=v_2$.
The same arguments can be used to show that there is at most a single
local maximum.

We are now able to start a detailed case by case discussion of the
asymptotic behavior (in the regime of validity of perturbation theory)
of a theory described by the Lagrangian (\ref{lag}) both in
the infrared and in the ultraviolet limit.

\section{Classification of flow patterns}
\underline{A) $\l>0$}\\[4pt]
When $\t\to-\infty$, $G$ becomes large and perturbation theory is no longer
valid.
So, if there is any fixed point within the range of validity of perturbation
theory, it must occur for $\t\to+\infty$, where $G\sim e^{-2\l \t}$.
If the $y$'s escape attraction by the ultraviolet fixed point or if there is
no ultraviolet fixed point, then the $y$'s go to infinity.
One might wonder whether $Y=gy$ could still remain finite. That this is not so
follows from the fact that the $y$'s reach infinity within {\it finite} $\t$:
\beq {dy\0 d\t}\sim const.\times y^3\quad {\rm implies}
                \quad -y^{-2}\sim const.\times (\t -const.)   \eeq
Thus perturbation theory breaks down.

If the starting point is in the region of
attraction of the local maximum of $U$, all $y^\a_{ij}$ will go to zero.
Near the fixed point, 
\beq 
   y^\a_{ij}\sim const.\times e^{-\t(D_{ii}+D_{jj})}\sim const.\times
     G^{(D_{ii}+D_{jj})/(2\l)}   \eeq
Then, at lowest order in $G$, the evolution of $v$ will be determined by
Eq. (\ref{bv}) at $y=0$.
A simple example for a non-trivial fixed point of $v$ is the case of just
one real scalar which is a singlet under the gauge group: At $y=0$
\beq G^{-1}{dv\0dt}=3v^2+\l v.   \eeq
There is an ultraviolet fixed point at $v_{\rm FP}=-\l/3=-(22c_g-4S_F)/9$,
which is approached according to
\beq v-v_{\rm FP}\sim const.\times e^{-2\l\t}\sim const.\times G.\eeq
In general $v-v_{\rm FP}$ will behave like some positive power of $G$.
Describing the evolution of couplings in
terms of some other coupling is just the idea of the
``reduction of coupling constants'' program \cite{red}.
The exponents of $G$ encountered in $y$ and $v-v_{\rm FP}$, which are
not necessarily integral, correspond precisely (modulo the fact
that different models were considered) to the non-integral
exponents found in \cite{red}.

On the other hand there are many theories without a fixed point for $v$,
due to the following argument (adapted from Ref. \cite{bd}):
A little calculation shows that at $y=0$
\beq G^{-1}{dv^{\a\a\g\g}\0dt}=
     (v^{\a\a\l\e}-6C_B^{\l\e}+{\l\0 2}\d^{\a\l})^2+2v^{\a\g\l\e}v^{\a\g\l\e}
     -12d_gQ_B+6d_gS_B(\l+2S_B-c_g)-{d_B\l^2\0 4},   \eql{bdtrick}
which is positive definite for many theories.
In such a case, or whenever we start outside the domain of attraction of an
ultraviolet fixed point, for large $\t$
\beq {dv\0d\t}\sim const.\times v^2,\quad v^{-1}\sim const.\times (const.-\t),
             \eeq
i.e. we leave the region of validity of perturbation theory within finite $\t$.

\ni\underline{B) $\l<0$}\\[4pt]
Here $G$ becomes large for $\t\to+\infty$, so that a perturbative fixed point
is possible only in the infrared limit $\t\to-\infty$.
In this case $y$ will always approach some infrared fixed point.
According to the discussion after Eq. (\ref{hess}), if
$N_{\rm flat}+$rank$(H)$ is equal to the dimension of $y$-space,
$y-y_{\rm FP}$ behaves like some exponential of $\t$ and therefore like a
positive power of $G$.
If this is not the case, it is possible that
\beq y-y_{\rm FP}\sim const.\times (const.-\t)^{-1/2}\sim
                      const.\times (\ln(G^{-1}))^{-1/2},\eeq
which is certainly a very slow approach to the fixed point.
If a fixed point for $v$ (at $y=y_{\rm FP}$) exists and the starting values
are in its domain of attraction, $v$ will also go to the fixed point.

As an example consider the case of $N$ chiral fermions in some complex
representation $R$ of the gauge group and as many fermions in the conjugate
representation $\bar R$,
together with a single scalar singlet.
By a biunitary transformation one can diagonalize the couplings, so that
$y_{i\bar j}=Z_{I(i)}\d_{i\bar j}$ (cf. Eq. (\ref{dec})),
where $I$ runs from 1 to $N$.
Then $U$ turns out to be of the form
\beq U=a(\sum_I|Z_I|^2)^2+b\sum_I|Z_I|^4-c\sum_I|Z_I|^2+d \eeq
with positive constants $a,b,c,d$.
$U$ is easily minimized with the result that at the fixed point all $|Z_I|^2$
are equal to some constant depending on $N$ and the dimension and
Casimir eigenvalue of $R$.
At $y=y_{\rm FP}$, $dv/d\t$ is positive for $v\to\pm\infty$ and negative for
$v=0$, i.e. it must have two zeroes.
The larger of the two values of $v$ for which $dv/d\t$ is zero corresponds
to a minimum of $U_v$.

\ni\underline{C) $\l=0$}\\[4pt]
The behaviour of $G$ is dictated by
\beq {1\0 8G^2}{dG \0 d\t}= -2c_g^2+c_gS_F+Q_F+Q_B-
       {1\0 2d_g}\Tr(C_F{y^\b}^\hc y^\b)+O(G),   \eql{G2}
where we have used $\l=0$, i.e. $S_B=22c_g-4S_F$, in order to eliminate $S_B$.
$y$ evolves according to Eq. (\ref{yuk}) with $D=6C_F$.

Let us first consider the ultraviolet limit $\t\to+\infty$:
If no Yukawa interacting fermionic singlets are present, $y=0$ is a local
maximum of $U$.
If there are fermionic singlets, or if the starting
configuration is outside the domain of attraction, $y$ will go to infinity
within finite $\t$.
Let us now assume that perturbation theory remains valid for $\t\to+\infty$.
Then $\lim_{\t\to+\infty}y=0$ and stability of $G$, determined by
Eq. (\ref{G2}) at $y=0$, implies $Q_B\le 2c_g^2$ and $S_F\le 2c_g$.
Reinserting the latter inequality into $\l=0$, we get $S_B\ge 14 c_g$.
Putting this into Eq. (\ref{bdtrick}), we see that $v^{\a\a\g\g}$ will go to
infinity within finite $\t$. We conclude that for $\l=0$ there is no
perturbative ultraviolet fixed point.
Of course, all this is again only true if there
is no exact fine tuning of the initial values which would allow the couplings
to stay in the minimum or in a saddle point of the potential.
Such a case is considered in Ref. \cite{shap}.

For $\t\to-\infty$, $y$ will go to some infrared fixed point.
The behavior of $G$ is determined by Eq. (\ref{G2})
with $\Tr(C_F{y^\b}^\hc y^\b)$ evaluated at the one--loop fixed point for $y$.
If $U$ has flat directions, $y_{\rm FP}$ depends continually on the
initial conditions.
If all flat directions correspond to symmetries of $U$, or if the fixed point
is at $U=0$, $\Tr(C_F{y^\b}^\hc y^\b)$ is nevertheless stable under small
changes in the initial conditions.
If the r.h.s. of Eq. (\ref{G2}) is negative, perturbation theory will break
down; if it is positive, $G=0$ will be an infrared fixed point, approached as
$G\sim (const.-t)^{-1/2}$.
If $U$ takes its minimum at vanishing $S_{ijkl}$, $I^{\a\b}$
and $M_{ij}$, then $\Tr(C_F{y^\b}^\hc y^\b)=6\Tr C_F^2=6d_gQ_F$ and
\beq {1\0 8G^2}{dG \0 d\t}= -2c_g^2+c_gS_F-2Q_F+Q_B+{\rm h.O.}   \eeq
The behavior of $v$ depends on whether there is a fixed point for $v$ at
$y=y_{\rm FP}$ and on the initial conditions.
Of course the theory is stable in the infrared limit only if $v$ approaches
such a fixed point.

The most interesting case occurs when the r.h.s. of Eq. (\ref{G2}),
evaluated at $y=y_{\rm FP}$, vanishes. Then
\bea {d_g\0 4G^2}{dG \0 d\t}&=&
       \Tr(C_F({y^\b}^\hc y^\b-{y^\b_{\rm FP}}^\hc y_{\rm FP}^\b)) +O(G)\nn\\
   &=&\Tr(C_F({\D y^\b}^\hc \D y^\b+{y^\b_{\rm FP}}^\hc\D y^\b+
               {\D y^\b}^\hc y_{\rm FP}^\b)) +O(G) , \eal{GG2}
where $\D y=y-y_{\rm FP}$,
so it is crucial for the further discussion {\it how} the fixed point
is approached.
If there are components of $y$ behaving like $(const.-\t)^{-1/2}$,
$G$ will run to $0$ or $\infty$.
If, however, $N_{\rm flat}+$rank$(H)$ is equal to the dimension of $y$-space,
all components of $\D y$ will tend to zero exponentially in $\t$.
Again the same type of behaviour has been encountered in the context of
``reduction of coupling constants'' \cite{osz}.
The r.h.s. of Eq. (\ref{GG2}) receives an exponential inhomogeneity;
for suitable (but not exactly fine-tuned) initial values $G$ remains
finite for $\t\to-\infty$ at the present order in perturbation theory.
In a next step one has to consider higher order corrections to the evolution
of $y$.
Using the vanishing of the first two orders of $\b(g)$, one gets
\beq {dy\0 d\t}={\6 U\0\6 y^\hc}
   +{2\0 gG}\b^{(2)}(Y)+O(G^2)\eeq
with $\b^{(2)}(Y)=gG^2f(y,v)$.
The ansatz $y=y_{\rm FP}+Gy^{(1)}+\ldots$ yields
\beq G{dy^{(1)}\0 d\t}=G(Hy^{(1)}+2f(y_{\rm FP},v_{\rm FP}))+{\rm h.O.}\eeq
Thus $y^{(1)}$ goes to $-2H^{-1}f(y_{\rm FP},v_{\rm FP})$ exponentially with
exactly the same rate as $\D y$ goes to zero.
In a similar way $v_{\rm FP}$ gets shifted by a term of the
order of $G$.
The value of $y^{(1)}$ is relevant for
the $G^4$--term (which is then dominant) in $\b(G)$.
The same analysis can be repeated until eventually $G$ starts to run like
some power of $\t$.
One would expect that this should always happen at some order in perturbation
theory. 
This is however not the case for supersymmetric theories,
due to the theorem of Ref. \cite{gmz} that in such theories the vanishing of
all beta functions at $N$ loop level implies vanishing of the gauge beta
function at $N+1$ loop level.

Summing up our results, we can distinguish the following four cases:
\begin{itemize}
\item Perturbation theory remains valid neither in the infrared nor in the
ultraviolet limit.
\item The theory is unstable in the infrared limit but shows asymptotic
freedom: For $t\to\infty$, $g$ behaves according to $g\sim t^{-1/2}$.
The Yukawa couplings go to zero as powers of $g$ with exponents greater than 1;
the behavior of the quartic scalar couplings is determined by a nontrivial
fixed point of $V/g^2$.
\item The theory is unstable in the ultraviolet limit. In the infrared
limit $g$ goes to zero, 
$Y/g$ and $V/g^2$ approach nontrivial fixed points.
\item There is no ultraviolet stable fixed point; in the infrared limit all
couplings approach a fixed point.
\end{itemize}
There is no theory with stable fixed points both in the infrared and in the
ultraviolet limit.


\begin{thebibliography}{11}\let\bib=\bibitem
\addtolength{\itemsep}{-2pt}
\bib{calc} T. P. Cheng, E. Eichten and L.-F. Li, \prd 6 (1972) 2973;\\
     M. E. Machacek and M. T. Vaughn, \npb 222 (1983) 83, \npb 236 (1984) 221
    and \npb 249 (1985) 70
\bib{zam} A. B. Zamolodchikov, JETP Lett. {\bf 43} (1986) 565
\bib{wz} D. J. Wallace and R. K. P. Zia, Phys. Lett. {\bf A48} (1974) 325
         and \aph 92 (1974) 142
\bib{tf} H. Skarke, \ijmp 9 (1994) 711
\bib{k} G. Kranner, doctorate thesis, Techn. Univ. Wien (1990)
\bib{kk} G. Kranner and W. Kummer, \plb 259 (1991) 84
\bib{k93} G. Kranner, preprint TUW--93--05
\bib{ref2} S. Hamidi, J. Patera and J. H. Schwarz, \plb 141 (1984) 349;\\
           S. Rajpoot and J. G. Taylor, \plb 147 (1984) 91;\\
           F. X. Dong, X. D. Jiang and X. D. Zhou, \jpa 19 (1986) 3863;\\
           X. D. Jiang and X. J. Zhou, \plb 197 (1987) 156
\bib{gran} P. Grandits, \mpl 9 (1994) 1093 and preprint TUW-93-00
\bib{red} W. Zimmermann, \cmp 97 (1985) 211;\\
    R. Oehme and W. Zimmermann, \cmp 97 (1985) 569
\bib{bd} M. B"ohm and A. Denner, \npb 282 (1987) 206
\bib{shap} S. D. Odintsov and I. L. Shapiro, JETP Lett. {\bf 49} (1989) 125
          and \mpl 4 (1989) 1479
\bib{osz} R. Oehme, K. Sibold and W. Zimmermann, \plb 147 (1984) 115
\bib{gmz} M. T. Grisaru, B. Milewski and D. Zanon, \plb 155 (1985) 357
\end{thebibliography}
\end{document}